\documentstyle[preprint,aps]{revtex}
\preprint{USM-TH-89}
\begin{document}
\title{Some remarks about gauge-invariant Yang-Mills fields}
\author{ Patricio Gaete \thanks{E-mail: pgaete@fis.utfsm.cl}}
\address{Departamento de F\'{\i}sica, Universidad T\'ecnica F.
Santa Maria, Casilla 110-V, Valparaiso, Chile}
\maketitle

\begin{abstract}
In order to eliminate gauge variant degrees of freedom we study
the way to introduce gauge invariant fields in pure non-Abelian
Yang-Mills theory. Our approach is based on the use of the
gauge-invariant but path-dependent variables formalism. It is
shown that for a special class of paths these fields coincide with
the usual ones in some definite gauges. The interquark potential
is discussed by exploiting the rich structure of the gluonic cloud
or dressing around static fermions.
\end{abstract}
\smallskip

PACS number(s): 12.20.Ds, 11.15.Tk

\section{INTRODUCTION}

The property of gauge invariance is at the origin of one of the
important advances of theoretical physics, that is, with the
introduction of non-Abelian gauge theories it was possible the
development of quantum chromodynamics (QCD) as the theory of
strong interactions, whose properties at short distances can be
calculated from perturbation theory. As is well known, for short
distances the property of asymptotic freedom is valid, and this
explains why perturbation theory can be used at sufficiently large
values of the momenta. But within this framework, we cannot
explain low energy phenomena such as the permanent confinement of
quarks and gluons. The analysis of this problem directly from the
QCD lagrangian is extremely difficult. The reason is that the
infrared divergences and gauge dependence make bound-state
equations very hard to approximate. Along the same line, we also
recall that the choice of the gauge has a strong influence on the
properties of the propagator\cite{Hatfield}. An illustrative
example on this subject arises when one considers the infrared
behavior of the fermion propagator in QCD2. In fact, it has been
emphasized that ambiguities appears with the choice of the
infrared regularization when a gauge-dependent fermion propagator
is treated in the $1/N$ approximation\cite{Schevc}. We also recall
that the studies of Caracciolo et al.\cite{Carac} were crucial in
the investigation of the ambiguity in the definition of the gluon
propagator. As a way to circumvent these difficulties there has
been a recent renewal of interest in formulations of QCD in which
gauge-invariant variables are explicitly constructed. By doing so
it has been possible to obtain a more deeper and illuminating view
on the description of charged particles \cite{Lavelle,Shabanov}.
It should be noted that the picture which emerges from these
studies is that quarks are dressed objects, where this dressing is
viewed as surrounding the quark with a cloud of gauge fields.

In previous work\cite{Gaete1,Gaete2,Gaete3} we have investigated
the gauge-invariant but path-dependent formalism in Abelian gauge
theories, and the intimately related question of gauge fixing. We
illustrated how the gauge fixing procedure corresponds, in this
formalism, to a path choice. We developed a path-dependent but
physical QED where a consistent quantization directly in the path
space was carried out. It is worthwhile remarking at this point
that the physical electron (dressed) is not the Lagrangian
fermion, which is neither gauge invariant nor associated with an
electric field. Instead, the physical electron is the Lagrangian
fermion accompanied with a non-local cloud of gauge fields.
Following this line of argument we reconsidered the calculation of
the interaction energy in pure QED and Maxwell-Chern-Simons gauge
theory, paying due attention to the structure of the fields that
surround the charges. In such a case subtleties related to the
problem of exhibiting explicitly the interaction energies were
illustrated. Subsequently, we have considered QED2 with associated
issues such as screening and confinement, producing computational
rules that have clear as well as simple
interpretations\cite{Gaete3}.

Inspired by these observations, in this Brief  Report we further
pursue the gauge-invariant but path-dependent variables formalism
by studying the extension to non-Abelian gauge fields. In Sec.II
the gauge-invariant variables are introduced. The general approach
is similar to that of other authors\cite{Schevc,Huang,Mandelstam},
paying due attention to the question of gauge choice. As an
application of the formalism, we consider the problem of
exhibiting explicitly the Coulombic interaction between pointlike
sources in a non-Abelian pure Yang-Mills theory (gluodynamics) in
Sec.III. As already expressed these variables are non-local and
require the introduction of strings to carry electric flux.

\section{GAUGE-INVARIANT FIELDS}
\subsection{The Poincar\'e gauge}

The aim of this section is to discuss the way to introduce
gauge-invariant fields in Yang-Mills theory. Towards such an end
we will begin by giving a brief account of some of the previous
work on Abelian gauge-invariant fields\cite{Gaete1} which is
relevant to the work at hand. We also recall that our line of
thought is to construct variables which are themselves unaltered
by a gauge transformation.  Accordingly, we consider the
gauge-invariant field
\begin{equation}
{\cal A}_{\mu }(x)=A_{\mu }(x)+\partial _{\mu }\left(
-\int_{C_{\xi x}}dz^{\nu }A_{\nu }(z)\right) ,  \label{bia}
\end{equation}
where the path integral is to be evaluated along some contour
$C_{\xi x\text{ }}$ connecting $\xi $ and $x$. Here $A_{\mu }$ is
the usual electromagnetic potential and, in principle, it is taken
in an arbitrary gauge. What we would like to stress is that by
choosing a spacelike path from the point $ \xi^k $ to $ \ x^k $,
on a fixed time slice, the expression (\ref{bia}) may be rewritten
as :
\begin{equation}
{\cal A}_0 \left( {x,\xi } \right) =  - \int\limits_0^1 {d\lambda
\left( {x - \xi } \right)} ^k F_{0k} \left( {\xi  + \lambda
\left( {x - \xi } \right)} \right), \label{bia1}
\end{equation}
\begin{equation}
{\cal A}_i \left( {x,\xi } \right) =  - \int\limits_0^1 {d\lambda
\lambda \left( {x - \xi } \right)} ^k F_{ki} \left( {\xi  + \lambda
\left( {x - \xi } \right)} \right), \label{bia2}
\end{equation}
where $\lambda $ $ (0$ $\leq \lambda $ $\leq 1)$ is the parameter
describing the contour $ z^k = \xi ^k + \lambda (x -\xi )^k $ with
$k=1,2,3$ ; and, as usual, $ F_{\mu \nu }$ stands for the field
strength tensor $ F_{\mu \nu }=\partial _\mu A_\nu-\partial _\nu
A_\mu $. Thus, as seen from (\ref{bia1}) and (\ref{bia2}), the
potentials ${\cal A}_\mu \left( {x,\xi } \right)$ are expressed in
a simple way in terms of the gauge field strengths $ F_{\mu\nu}$.
In passing we note that the field strengths are gauge-invariant
observables in the Abelian case, in contrast to the non-Abelian
theories where the field strengths are gauge covariant rather than
invariant. This raises the question of how to construct
gauge-invariant fields in the non-Abelian case. It is the purpose
of the present section to provide such construction. Before going
into details, we would like to add here that the expressions
(\ref{bia1}) and (\ref{bia2}) coincide with the Poincar\'e gauge
conditions\cite{Pimentel} defined by
\begin{equation}
\left( {x - \xi } \right)^i A_i \left( x \right) = 0 , \label{pon1}
\end{equation}
\begin{equation}
A_0 \left( x \right) = \int\limits_0^1 {d\lambda \left( {x - \xi }
\right)^i } \Pi ^i \left( {\lambda x} \right), \label{pon2}
\end{equation}
where $\Pi ^i$ is the conjugate momentum to $ A^i$.

Our immediate undertaking is to extend the above derivation to the
non-Abelian case. The first step in this direction is to define
the following fields
\begin{eqnarray}
A_0 \left( {x,\xi } \right) = A_0 \left( x \right) - \partial _0
\int\limits_0^1 {d\lambda \left( {x - \xi } \right)} ^k A_k \left(
{\xi  + \lambda \left( {x - \xi } \right)} \right) + \nonumber \\
+ ig\int\limits_0^1 {d\lambda } \left( {x - \xi } \right)^k \left[
{A_0 \left( {\xi  + \lambda \left( {x - \xi } \right)} \right),A_k
\left( {\xi  + \lambda \left( {x - \xi } \right)} \right)}
\right], \label{pab1}
\end{eqnarray}
\begin{eqnarray}
A_i \left( {x,\xi } \right) = A_i \left( x \right) - \partial _i
\int\limits_0^1 {d\lambda \left( {x - \xi } \right)} ^k A_k \left(
{\xi  + \lambda \left( {x - \xi } \right)} \right) + \nonumber \\
+ ig\int\limits_0^1 {d\lambda }\lambda \left( {x - \xi } \right)^k
\left[ {A_i \left( {\xi  + \lambda \left( {x - \xi } \right)}
\right),A_k \left( {\xi  + \lambda \left( {x - \xi } \right)}
\right)} \right]. \label{pab2}
\end{eqnarray}
To transform (\ref{pab2}) we use
\begin{equation}
G_{\mu\nu} \left( x \right) = \partial _\mu A_\nu \left( x\right)-
\partial _\mu A_\nu \left( x \right) - ig\left[ {A_\mu \left( x
\right),A_\nu \left( x \right)} \right] , \label{fer}
\end{equation}
which allows us to rewrite (\ref{pab2}) in the form
\begin{eqnarray}
A_i \left( {x,\xi } \right) = A_i \left( x \right) -
\int\limits_0^1 {d\lambda \frac{d}{{d\lambda }}} \left( {\lambda
A_i \left( {\xi  + \lambda \left( {x - \xi } \right)} \right)}
\right) +  \nonumber \\ - \int\limits_0^1 {d\lambda \lambda \left(
{x - \xi } \right)} ^k G_{ik} \left( {\xi  + \lambda \left( {x -
\xi } \right)} \right) . \label{fer1}
\end{eqnarray}
The first integral in (\ref{fer1}) is found immediately, and it
exactly cancels the first term of (\ref{fer1}), so that
\begin{equation}
A_i \left( {x,\xi } \right) = \int\limits_0^1 {d\lambda \lambda
\left( {x - \xi } \right)^k G_{ki} \left( {\xi  + \lambda \left(
{x - \xi } \right)} \right)} . \label{for1}
\end{equation}
Proceeding in a similar manner, one gets the following expression
for (\ref{pab1}),
\begin{equation}
A_0 \left( {x,\xi } \right) =  - \int\limits_0^1 {d\lambda \left(
{x - \xi } \right)} ^k G_{0k} \left( {\xi  + \lambda \left( {x -
\xi } \right)} \right) . \label{for2}
\end{equation}
Except for the substitution of $F_{\mu\nu}$ to $ G_{\mu \nu }$ we
see that the non-Abelian inversion formula for the fields
(\ref{pab1}) and (\ref{pab2}) coincides with (\ref{bia1}) and
(\ref{bia2}). In fact, such expressions are the non-Abelian
inversion formula in the Poincar\'e gauge by application of the
Poincar\'e lemma to the field strength two-form\cite{Galvao}.
However, as outlined above, the fields (\ref{for1}) and
(\ref{for2}) are not gauge invariant. Thus, as a possible way to
introduce gauge invariant fields, we perform the gauge
transformations
\begin{equation}
A_{\mu} \left( x \right) \to A_{\mu}^\Lambda  \left( x \right) =
\Lambda ^{ - 1} \left( x \right)\left( {A_{\mu} \left( x \right) -
\frac{i}{g}\partial _{\mu} } \right)\Lambda \left( x \right),
\label{cali1}
\end{equation}
\begin{equation}
G_{\mu \nu } \left( x \right) \to G_{\mu \nu }^\Lambda  \left( x
\right) = \Lambda \left( x \right)G_{\mu \nu } \left( x
\right)\Lambda ^{ - 1} \left( x \right) , \label{cali2}
\end{equation}
with the gauge transformation $ \Lambda \left( x \right) =
U^{\dag} \left( {x,\xi } \right)$. The operator $ U(x,\xi )$ is
defined by the P-ordered exponential
\begin{equation}
U\left( {x,\xi } \right) = P\exp \left( { - ig\int\limits_\xi ^x
{dz^i A_i \left( z \right)} } \right) , \label{wil}
\end{equation}
with the integration path corresponding to the spacelike straight
line which connects $\xi$ to $x$. Let us also recall here that the
P-ordered exponential transforms under a gauge transformation
$\Lambda \left( x \right)$ as
\begin{equation}
U\left( {x,\xi } \right) \to U^\Lambda  \left( {x,\xi } \right) =
\Lambda \left( x \right)U\left( {x,\xi } \right)\Lambda ^{ - 1}
\left( \xi  \right) . \label{wil1}
\end{equation}
As a consequence $ G_{0k} \left( {x,\xi } \right)$ and $ G_{ki}
\left( {x,\xi } \right)$ take the form
\begin{equation}
G_{0k} \left( {x,\xi } \right) \to  G_{0k}^U \left( {x,\xi }
\right) \equiv {\cal G}_{0k} \left( {x,\xi } \right) = U^{\dag}
\left( {x,\xi } \right)G_{ok} \left( {x,\xi } \right)U\left(
{x,\xi } \right), \label{pau1}
\end{equation}
\begin{equation}
G_{ki} \left( {x,\xi } \right) \to  G_{ki}^U \left( {x,\xi }
\right) \equiv {\cal G}_{ki} \left( {x,\xi } \right) = U^{\dag}
\left( {x,\xi } \right)G_{ki} \left( {x,\xi } \right)U\left(
{x,\xi } \right), \label{pau2}
\end{equation}
and similarly the fields $ A_0 \left( {x,\xi } \right)$ and $A_i
\left( {x,\xi } \right)$ will transform into $ {\cal A}_0 \left(
{x,\xi } \right)$ and $ {\cal A}_i \left( {x,\xi } \right)$
respectively. The point we wish to emphasize, however, is that
under the local gauge transformations the fields (\ref{pau1}) and
(\ref{pau2}) transform globally. To see this, let us study how a
gauge transformation affects the fields (\ref{pau1}) and
(\ref{pau2}), that is,
\begin{equation}
{\cal G}_{0k} \left( {x,\xi } \right) \to {\cal G}_{0k}^\Lambda
\left( {x,\xi } \right) = \left( {U^{\dag}  \left( {x,\xi }
\right)} \right)^\Lambda {\cal G}_{0k}^\Lambda  \left( {x,\xi }
\right)U^\Lambda \left( {x,\xi } \right) , \label{pau3}
\end{equation}
\begin{equation}
{\cal G}_{ki} \left( {x,\xi } \right) \to {\cal G}_{ki}^\Lambda
\left( {x,\xi } \right) = \left( {U^{\dag}  \left( {x,\xi }
\right)} \right)^\Lambda {\cal G}_{ki}^\Lambda  \left( {x,\xi }
\right)U^\Lambda \left( {x,\xi } \right) . \label{pau4}
\end{equation}
Using the expressions (\ref{cali1}) and (\ref{wil1}) a short
calculation yields
\begin{equation}
{\cal G}_{0k}^\Lambda  \left( {x,\xi } \right) = \Lambda \left(
\xi \right){\cal G}_{0k} \left( {x,\xi } \right)\Lambda ^{ - 1}
\left( \xi \right) , \label{pau5}
\end{equation}
\begin{equation}
{\cal G}_{ki}^\Lambda  \left( {x,\xi } \right) = \Lambda \left(
\xi \right){\cal G}_{ki} \left( {x,\xi } \right)\Lambda ^{ - 1}
\left( \xi \right) . \label{pau6}
\end{equation}
The above result explicitly shows that the fields (\ref{pau1}) and
(\ref{pau2}), under local gauge transformations, transform only
globally. This allows us to say that the gauge invariance of
(\ref{pau5}) and (\ref{pau6}) is retrieved when $ \Lambda
\left({\xi } \right) \to 1$, which is achieved by letting to point
$\xi$ go to infinity. In a similar manner we find that the gauge
invariance of the fields ${\cal A}_0 \left( {x,\xi } \right)$ and
$ {\cal A}_i \left( {x,\xi } \right)$ is restored when $ \Lambda
\left( {\xi  } \right) \to 1$. As a consequence of this one can
construct Yang-Mills gauge invariant, not merely covariant,
fields. On the other hand, if we had selected from the beginning
the path in the form of a spacelike straight line, where the
reference point $\xi$ is in spatial infinity, we would arrive at
an expression for ${\cal A}_\mu$ that coincides with the axial
gauge. In order to show this more clearly it is instructive to
repeat the above derivation for the infinite reference point case.

\subsection{The Axial gauge}

First let us discuss the Abelian case. For this purpose we start
by considering the gauge invariant field given by
\begin{equation}
{\cal A}_\mu  \left( x \right) = A_\mu  \left( x \right) +
\partial _\mu \left( { - \int\limits_{ - \infty }^0 {A_\sigma
\left( z \right)\frac{{\partial z^\sigma  }}{{\partial \xi }}d\xi
} } \right) . \label{axi1}
\end{equation}
As already expressed the reference point $\xi$ is in infinity and
$ C_{\xi x}$ is a spacelike path or contour that to join the
points $\xi$ and $x$. Now, $ z^\mu  \left( {x,\xi } \right)$ are
four arbitrary single-valued differential functions that satisfy
\begin{equation}
z^\mu  \left( {x,0 } \right) = x^\mu , \label{conda}
\end{equation}
\begin{equation}
z^\mu  \left( {x,\xi  \to  - \infty } \right)= \infty (spatial).
\label{condb}
\end{equation}
Just as for the Poincar\'e gauge case, after a brief calculation,
from (\ref{axi1}) one gets
\begin{equation}
{\cal A}_\mu  \left( x \right) = \int\limits_{ - \infty }^x
{F_{\nu \sigma } \left( z \right)} \frac{{\partial z^\sigma
}}{{\partial x^\mu  }}\frac{{\partial z^\nu  }}{{\partial \xi
}}d\xi .  \label{axi2}
\end{equation}
As before, the potentials ${\cal A}_\mu $ have the property that
they can be expressed in terms of the field strength $ F_{
\mu\nu}$. To illustrate a practical use of expression
(\ref{axi2}), we can write $ z^\mu \left( {x,\xi } \right) = x^\mu
+ \xi n^{\mu} $ with the vector $n^{\mu}=(0,0,0,1)$  and $ -
\infty < \xi  \le 0 $, in which case ${\cal A}_\mu  \left( x
\right)$ is said to be in the axial gauge.

The next step is to extend this analysis to the non-Abelian case.
Basically, we follow the same procedure as we developed in the
Poincar\'e gauge case. According to this idea one writes
\begin{equation}
A_\mu  \left( {x,\xi } \right) = A_\mu  \left( x \right) +
\partial _\mu  \left( { - \int\limits_{ - \infty }^0 {A_\sigma
\left( z \right)\frac{{\partial z^\sigma  }}{{\partial \xi }}} }
\right) - ig\int\limits_{ - \infty }^0 {d\xi \frac{{\partial
z^\sigma  }}{{\partial \xi }}} \frac{{\partial z^\rho }}{{\partial
x^\mu  }}\left[ {A_\sigma  \left( z \right),A_\rho \left( z
\right)} \right] , \label{axi3}
\end{equation}
which by its turn is rewritten as
\begin{equation}
A_\mu  \left( {x,\xi } \right) =  - \int\limits_{ - \infty }^0
{d\xi \frac{{\partial A_\sigma  }}{{\partial z^\rho  }}}
\frac{{\partial z^\rho  }}{{\partial x^\mu  }}\frac{{\partial
z^\sigma  }}{{\partial \xi }} + \int\limits_{ - \infty }^0 {d\xi }
\frac{{\partial A_\sigma  }}{{\partial \xi }}\frac{{\partial
z^\sigma  }}{{\partial x^\mu  }} - ig\int\limits_{ - \infty }^0
{d\xi \frac{{\partial z^\sigma  }}{{\partial \xi }}}
\frac{{\partial z^\rho  }}{{\partial x^\mu  }}\left[ {A_\sigma
\left( z \right),A_\rho  \left( z \right)} \right] . \label{axi4}
\end{equation}
By means the definition for the field strength in terms of the
vector potential
\begin{equation}
G_{\nu \rho }  = \partial _\sigma  A_\rho   - \partial _\rho
A_\sigma   - ig\left[ {A_\sigma  ,A_\rho  } \right] ,\label{axi5}
\end{equation}
the expression (\ref{axi4}) then becomes
\begin{equation}
A_\mu  \left( {x,\xi } \right) = \int\limits_{ - \infty }^0
{dz^\nu  } G_{\nu \sigma } \left( {z\left( \xi  \right)}
\right)\frac{{\partial z^\sigma  }}{{\partial x^\mu  }}.
\label{axi6}
\end{equation}
In other words, one obtains a non-Abelian inversion formula which
uniquely expresses the potentials $A_\mu  \left( {x,\xi } \right)$
in terms of the gauge field strengths $G_{\mu\nu}$, like in the
Poincar\'e gauge case. We now proceed to perform the gauge
transformation (\ref{cali1}) and (\ref{cali2}) with the operator
(\ref{wil}), and now with the spacelike path of integration
running from infinity to $x$. Proceeding in analogy to the
Poincar\'e gauge case and after some manipulations we find that $
{\cal G}_{\mu\nu}\left( {x,\xi } \right)\equiv
G^\Lambda_{\mu\nu}\left( {x,\xi } \right)$ and ${\cal A}_\mu\left(
{x,\xi } \right)\equiv A^\Lambda_\mu\left( {x,\xi} \right)$ are
gauge invariant.

\section{INTERQUARK ENERGY}
The goal of this section is to implement the above general
considerations in a concrete application. So we proceed to
calculate the interaction energy between external probe sources in
a non-Abelian pure Yang-Mills theory. This calculation will be
carry out with the help of the previously discussed string
variables. Before considering explicitly the interquark energy, we
shall first reexamine the canonical quantization for the
Yang-Mills field coupled to an external source $ J^0$ from the
viewpoint of hamiltonian dynamics. We start from the Lagrangian
density
\begin{equation}
{\cal L} =  - \frac{1}{4}Tr\left( {F_{\mu \nu } F^{\mu \nu } }
\right)-A_0^aJ^0 = - \frac{1}{4}F_{\mu \nu }^a F^{a\mu \nu }
-A_0^aJ^0. \label{rico1}
\end{equation}
Here $ A_\mu  \left( x \right) = A_\mu ^a \left( x \right)T^a $,
where $T^a$ is a hermitian representation of the semi-simple and
compact gauge group; and $ F_{\mu \nu }^a  = \partial _\mu  A_\nu
^a - \partial _\nu  A_\mu ^a  + gf^{abc} A_\mu ^b A_\nu ^c$, with
$f^{abc}$ the structure constants of the gauge group. The Dirac
procedure \cite{Dirac2} as applied to (\ref{rico1}) is
straightforward. The canonical momenta are $ \Pi ^{a\mu } = -
F^{a0\mu }$, which results in the usual primary constraint $ \Pi
_0^a  = 0$, and $ \Pi ^{ai}  = F^{ai0}$. This allows us to write
the following canonical Hamiltonian:
\begin{equation}
H_c  = \int {d^3 x} \left( { - \frac{1}{2}\Pi _i^a \Pi _a^i  + \Pi
_i^a \partial ^i A_a^0  + \frac{1}{4}F_{ij}^a F^{aij}  - gf_{abc}
\Pi _i^a A_b^0 A_c^i }+A_0^aJ^0 \right) . \label{rico2}
\end{equation}
The secondary constraint generated by the time preservation of the
primary constraint is now
\begin{equation}
\Omega _a^{\left( 1 \right)} \left( x \right) = \partial _i \Pi
_a^i  + gf_{abc} A_b^i \Pi _i^c-J^0  \approx 0 . \label{secon}
\end{equation}
It is easy to check that there are no more constraints in the
theory,  and that both constraints are first class. The
corresponding total (first class) Hamiltonian that generates the
time evolution of the dynamical variables is given by
\begin{equation}
H = H_c  + \int {d^3 x\left( {c_0 \left( x \right) + c_1 \left( x
\right)\Omega _a^{\left( 1 \right)} \left( x \right)} \right)} ,
\label{rico3}
\end{equation}
where $c_0$ and $c_1$ are arbitrary functions. Since $ \Pi_0^a
\approx 0$ for all time and $ \dot{A}_0^a \left( x \right) =
\left[ {A_0^a \left( x \right),H} \right] = c_0 \left( x \right)$,
which is completely arbitrary, we discard $ A_0^a \left( x
\right)$ and $ \Pi _a^0 \left( x \right)$ since they add nothing
to the description of the system. The Hamiltonian then takes the
form
\begin{equation}
H = \int {d^3 } x\left( { - \frac{1}{2}\Pi _i^a \Pi _a^i  +
\frac{1}{4}F_{ij}^a F^{aij}  + c^a \left( x \right)\left(
{\partial ^i \Pi _i^a  + gf_{abc} A_b^i \Pi _i^c-J^0 } \right)}
\right) , \label{rico4}
\end{equation}
where $ c^a \left( x \right) = c_1 \left( x \right) - A_0^a \left(
x \right)$.

Therefore we have one first class constraint $ \Omega _a^{\left( 1
\right)} \left( x \right)$ , which appears at the secondary level.
In order to break the gauge freedom of the theory, it is necessary
to impose one constraint such that the full set of constraints
becomes second class. From the previous section, we choose
\begin{equation}
\Omega _a^{\left( 2 \right)} \left( x \right) = \int\limits_0^1
{d\lambda } \left( {x - \xi } \right)^k A_k^{\left( a \right)}
\left( {\xi  + \lambda \left( {x - \xi } \right)} \right) \approx
0   . \label{poinca1}
\end{equation}
There is no essential loss of generality if we restrict our
considerations to $ \xi^k=0$; accordingly, (\ref{poinca1}) becomes
\begin{equation}
\Omega _a^{(2)} \left( x \right) = \int\limits_0^1 {d\lambda } x^k
A_k^a \left( {\lambda x} \right) \approx 0 . \label{poinca2}
\end{equation}

Standard techniques for constrained systems then lead to the
following Dirac brackets
\begin{equation}
\left\{ {A_i^a \left( x \right),A_b^j \left( y \right)} \right\}^
*   = 0 = \left\{ {\Pi _i^a \left( x \right),\Pi _b^j \left( y
\right)} \right\}^ * , \label{dirb1}
\end{equation}

\begin{equation}
\left\{ {A_i^a \left( x \right),\Pi ^{bj} \left( y \right)}
\right\}^ *   = \delta ^{ab} \delta _i^j \delta ^{(3)} \left( {x -
y} \right) - \int\limits_0^1 {d\lambda } \left( {\delta ^{ab}
\frac{\partial }{{\partial x^i }} - gf^{abc} A_i^c \left( x
\right)} \right)x^j \delta ^{(3)} \left( {\lambda x - y} \right) .
\label{dirb2}
\end{equation}
Also we mention that equivalent Dirac brackets were calculated
independently in reference \cite{Poland}. This completes our brief
review of the canonical quantization for the Yang-Mills field.

Now we move on to compute the energy of the external field of
static charges where a fermion is localized at $ {\bf y^\prime}$
and an antifermion at $ {\bf y}$. In order to accomplish this
purpose we will calculate the expectation value of the energy
operator $ H$ in the physical state $ |\Omega\rangle$, which will
denote by ${ \langle H \rangle_ \Omega}$ . Using Eq.(\ref{rico4})
we see that ${ \langle H \rangle_ \Omega}$ reads
\begin{equation}
\left\langle H \right\rangle _\Omega   = tr\left\langle \Omega
\right|\int {d^3 x} \left( { - \frac{1}{2}\Pi _i^a \Pi ^{ia}  +
\frac{1}{4}F_{ij}^a F^{aij} } \right)\left| \Omega  \right\rangle.
\label{sort1}
\end{equation}
Since the fermions are taken to be infinitely massive (static),
this can be further simplified as
\begin{equation}
\left\langle H \right\rangle _\Omega   = tr\left\langle \Omega
\right|\int {d^3 x} \left( { - \frac{1}{2}\Pi _i^a \left( x
\right)\Pi ^{ia} \left( x \right)} \right)\left| \Omega
\right\rangle . \label{sort2}
\end{equation}
It is now important to notice that, as was first established by
Dirac \cite {Dirac}, the physical states $|\Omega\rangle$
correspond to the gauge invariant ones. It is worth recalling at
this stage that in the Abelian case $|\Omega\rangle$ may be
written as \cite{Gaete2}
\begin{equation}
\left| \Omega  \right\rangle  = \overline \psi \left( {\bf y}
\right)\exp \left( {ie\int\limits_{\bf y^\prime}^{\bf y} {dz^i A_i
\left( z \right)} } \right)\psi \left( {\bf y^\prime}
\right)\left| 0 \right\rangle , \label{sort3}
\end{equation}
where $|0\rangle$ is the physical vacuum state and the line
integral appearing in the above expression is along a spacelike
path from the point at $ \bf {y^\prime}$ to $\bf y$, on a fixed
time slice. As a result of this the fermion fields are now dressed
by a cloud of gauge fields. As before, the above expression must
be extended since we are dealing with non-Abelian fields. Thus, on
the basis of the discussion in the previous section, we can write
a state which has a fermion at $ \bf {y^\prime}$ and an
antifermion at $\bf y$ as
\begin{equation}
\left| \Omega  \right\rangle  = \overline \psi \left( {\bf y}
\right)P \exp \left( {ig\int\limits_{\bf y^\prime}^{\bf y} {dz^i
A_i \left( z \right)} } \right)\psi \left( {\bf y^\prime}
\right)\left| 0 \right\rangle. \label{sort4}
\end{equation}
As before, the line integral is along a spacelike path on a fixed
time slice, and $|0\rangle$ is the physical vacuum state.

The above analysis give us an opportunity to compare our work with
the standard Wilson loop procedure \cite{Peskin}, to make sure
that the known results are recovered from the general expression
(\ref{sort4}) in the weak coupling limit. In effect, due to
asymptotic freedom, the short distance behavior of the interquark
potential is determined by perturbation theory. According to this,
at weak coupling, one can expand
\begin{equation}
P\exp \left( {ig\int\limits_{\bf y^\prime}^{\bf y} {dz^i } A_i
\left( z \right)} \right) = P \left( {1 + ig\int\limits_{\bf
y^\prime}^{\bf y} {dz^i } A_i^a \left( z \right)T^a  + ....}
\right) .\label{sort5}
\end{equation}
This implies that, at lowest order in g, the non-Abelian
generalization of the dressing framework is the same as in the
Abelian theory. In other terms, this means that at short distances
we get the Coulomb potential with $ g^2$ multiplied by a group
factor, as we will now show.

It is appropriate to start first by reconsidering how the gauge
invariant state (dressed) represent charged particles with a
static field electric on a line or, more precisely, on a tube. Let
$ |E\rangle$ be an eigenvector of the electric field operator $
E_i^a(x)$, with eigenvalue ${\varepsilon_i^a}(x)$:
\begin{equation}
E_i^a \left( x \right)\left| E \right\rangle  = \varepsilon _i^a
\left( x \right)\left| E \right\rangle .\label{nic1}
\end{equation}
We then focus our attention towards examining the state
\begin{equation}
U\left( {\bf y^\prime}  ,\bf y \right) \left| E \right\rangle
\equiv \overline \psi \left( {\bf y} \right)\left( {1 +
ig\int\limits_{\bf y^\prime}^{\bf y} {dz^i } A_i^c \left( z
\right)T^c} \right)\psi \left( {\bf y^\prime} \right)\left| E
\right\rangle. \label{nic2}
\end{equation}
Moreover, it follows from (\ref{dirb2}) and (\ref{nic1}) that
\begin{equation}
E_i^a \left( x \right)U\left( {\bf y^\prime}  ,\bf y \right)\left|
E \right\rangle  = \left( {\varepsilon _i^a \left( x \right) +
g\int\limits_{\bf y^\prime }^{\bf y} {dz_i \delta ^{(3)} \left( {x
- z} \right)T^a } } \right)U\left( {\bf y^\prime}  ,\bf y
\right)\left| E \right\rangle .\label{nic3}
\end{equation}
This means that $U\left( {\bf y^\prime}  ,\bf y \right) \left| E
\right\rangle$ is another eigenvector of $E_i^a(x)$ with
eigenvalue of $ {\varepsilon _i^a \left( x \right) +
g\int\limits_{\bf y^\prime }^{\bf y} {dz_i \delta ^{(3)} \left( {x
- z} \right)T^a } }$. As in \cite{Gaete2}, by employing
(\ref{sort2})and (\ref{nic3}) we then evaluate the energy in the
presence of the static charges. Hence we once again obtain
\begin{equation}
\left\langle H \right\rangle _\Omega   = \left\langle H
\right\rangle _0  + \frac{1}{2}g^2 trT^a T^a \int\limits_{\bf
y^\prime }^{\bf y} {dz^i } \int\limits_{\bf y^\prime  }^{\bf y} {dz_i^\prime
} \delta ^{(3)} \left( {z - z^\prime  } \right), \label{nic4}
\end{equation}
where $ \left\langle H \right\rangle _0 = \langle 0|H|0\rangle $.
Recalling that the integrals over $z^i$ and $z_i^\prime$ are zero
except on the contour of integration, we obtain the following
interaction energy
\begin{equation}
V = \frac{1}{2}g^2 k(trT^a T^a )|\bf y - {\bf y^\prime} |, \label{nic5}
\end{equation}
where $k=\delta^{(2)} (0)$. Writing the purely group theoretic
factor $trT^aT^a=C_2(F)$, (\ref{nic5}) can be further expressed as
\begin{equation}
V =   \frac{1}{2}g^2 C_2(F) k|\bf y - {\bf y^\prime} |.
\label{nic6}
\end{equation}
This expression may look peculiar, but it is nothing but the
familiar Coulomb energy as was discussed in \cite{Gaete2}. Here,
however, we call the attention to the fact that, as in the Abelian
case, the term$ \ \frac{{g^2 }}{2}\int {d^3 x\left(
{\int\limits_{\bf y^\prime  }^{\bf y} {dz_i \delta ^{(3)} \left(
{x - z} \right)} } \right)} ^2$ reproduces exactly the Coulomb
energy. In this way one obtains the standard result for the
interquark potential
\begin{equation}
V = - \frac{1}{{4\pi }}g^2 C_2(F) \frac{1}{{|{\bf y} - {\bf
y^\prime} |}}. \label{go1}
\end{equation}
We further note that, as was explained in \cite{Gaete2}, a
modified form for the state (\ref{sort4}) in the Poincar\'e gauge
is equivalent to the Coulomb gauge, which, too, leads to the
expression (\ref{go1}).

As a final comment, notice that the results of this section hinge
on the constraints structure of the theory we have discussed. Thus
it seems straightforward to extend to the $g^4$ order the
calculation that we have developed. We expect to report on
progress along these lines soon.

\section{ACKNOWLEDGMENTS}

The author would like to thank I. Schmidt for reading the
manuscript and for his support. Work supported in part by Fondecyt
(Chile) grant 1000710, and by a C\'atedra Presidencial en Ciencias
(Chile).

\end{document}